\documentclass[a4paper]{article}
\usepackage{graphicx}     
\usepackage{amsmath,amsfonts}
\usepackage{booktabs}
\usepackage[pagebackref=false]{hyperref}
\usepackage{subfig}
\usepackage{pgfplots}

\begin{document}

\title{Sparse grid high-order ADI scheme for option pricing in stochastic volatility models}

\author{Bertram D{\"u}ring\thanks{Department of Mathematics,
    University of Sussex, Pevensey II, Brighton, BN1 9QH, United
    Kingdom, \texttt{bd80@sussex.ac.uk}} \and Christian Hendricks\thanks{Chair of Applied Mathematics / Numerical Analysis,
  Bergische Universit{\"a}t Wuppertal, Gau\ss stra\ss e 20, 42097 Wuppertal,
Germany, \texttt{hendricks@uni-wuppertal.de}} \and James Miles\thanks{Department of Mathematics,
    University of Sussex, Pevensey II, Brighton, BN1 9QH, United
    Kingdom, \texttt{james.miles@sussex.ac.uk}}}

\maketitle
\begin{abstract}
\noindent We present a sparse grid high-order alternating direction implicit
(ADI) scheme for option pricing in stochastic volatility models.
The scheme is second-order in time and fourth-order in space.
Numerical experiments confirm the computational efficiency gains
achieved by the sparse grid combination technique.
\end{abstract} 

\section{Introduction}
\noindent Stochastic volatility models such as
the Heston model \cite{Hes93}
have become one of the standard approaches in financial option pricing.
For some stochastic volatility models and under additional restrictions, closed-form
solutions can be obtained by Fourier methods (e.g.\
  \cite{Hes93}, \cite{Due09}). 
Another approach is to derive approximate analytic expressions, see e.g.\
\cite{BeGoMi10} and the literature cited therein.
In general, however, ---even in the Heston model \cite{Hes93}  when the
parameters in it are non constant--- the partial differential
equations arising from stochastic volatility models have
to be solved numerically.

In the mathematical literature, there are many papers on numerical
methods for option pricing, mostly addressing the one-dimensional case of a single
risk factor and using standard, second order finite difference
methods (see, e.g., \cite{TavRan00} and the references therein). More
recently, high-order finite difference schemes (fourth order in space)
were proposed \cite{Gupta,Rigal99,SpotzCarey} that use a compact stencil (three points in space). In
the option pricing context, see e.g.\ \cite{DuFoJu04,DuFoJu03,LiaKha09}.

There are less works considering numerical methods for option
pricing in stochastic volatility models, i.e., for two spatial
dimensions. Finite difference approaches that are used are
often standard, second-order methods, e.g. in \cite{IkoToi07} where different efficient
methods for solving the American 
option pricing problem for the Heston model are proposed. In
\cite{DuFo12} a high-order compact finite difference scheme for option
pricing in the Heston model is derived and this approach is extended
to non-uniform grids in \cite{DuFoHe14}.
Other approaches include finite element-finite
volume \cite{ZvFoVe98}, multigrid \cite{ClaPar99}, sparse wavelet
\cite{HiMaSc05}, FFT-based \cite{Oosterlee} or spectral methods \cite{ZhuKop10}.

The classical alternating direction implicit (ADI) method, introduced by
Peaceman and Rachford \cite{PeacmanRachford95}, Douglas \cite{Douglas,
Gunn}, Fairweather and Mitchell \cite{Fairweather},
is a very powerful method that is especially
useful for solving parabolic equations ({\em without\/} mixed derivative
terms) on rectangular domains. Beam
and Warming \cite{BeamWarming}, however, have shown that no simple 
ADI scheme involving only discrete solutions at time levels $n$ and $n+1$ can be
second-order accurate in time in the presence of mixed derivatives. To overcome this limitation,
unconditionally stable ADI schemes which are second order in time have been proposed by Hundsdorfer and Verwer \cite{Hund02, Verwer} and 
more recently by in't Hout and Welfert \cite{HouWel07}.
These schemes are second-order accurate in time and space.
In \cite{HouFou10} different second-order ADI schemes of this type are
applied to the Heston model.
In \cite{DuFoRi13} this approach is combined with different
high-order discretisations in space, using high-order compact schemes
for two-dimensional
convection-diffusion problems {\em with mixed derivatives and constant
coefficients}.
In \cite{HeEhGu15} this approach is combined with sparse grids and
applied to multi-dimensional diffusion equations, again {\em with constant
coefficients}. 
Building on the ideas in \cite{Hund02, Verwer,DuFoRi13}, a high-order  (second-order accurate in time and fourth-order accurate in space) 
ADI method for option pricing in stochastic volatility models which 
involve the solution of two-dimensional convection-diffusion equations {\em with
  mixed derivative terms and space-dependent coefficients\/} is
derived in \cite{DM16}.
 
In this chapter we combine the approaches from \cite{HeEhGu15} and
\cite{DM16}, to obtain a {\em sparse grid high-order ADI scheme\/} for option
pricing in stochastic volatility models. In the next section we recall
stochastic volatility models for option pricing and the related
convection-diffusion partial differential equations. Section~\ref{sec:HV} 
is devoted to the Hundsdorfer-Verwer ADI splitting in time. The
spatial discretisation is introduced in
Section~\ref{sec:HOC} for the implicit steps, and in
Section~\ref{sec:HOexpl} for the explicit steps. The solution of the
resulting scheme and the discretisation of boundary conditions are discussed in
Sections~\ref{sec:HOsolve} and \ref{sec:bc}.
The sparse grid combination technique is explained in Section~\ref{sec:ct}.
We present numerical convergence results in
Section~\ref{sec:num}.

\section{Stochastic volatility models}

\noindent We consider the following class of stochastic volatility models: assume that asset spot price $0 \leq S(t)< \infty$ and variance $0 \leq \sigma(t)< \infty$ follow two stochastic diffusive processes for $t\in[0,T]$,
\begin{subequations}
\label{eq:SVmodels}
\begin{align}
dS(t)&=\mu S(t)dt + \sqrt{\sigma (t)} S(t) dW^{(1)}(t), \\
d\sigma(t)&=\kappa(\sigma(t))^\alpha(\theta-\sigma(t))dt + v(\sigma(t))^\beta dW^{(2)}(t) ,
\end{align}
\end{subequations}
which are characterised by two Brownian motions, $dW^{(1)}(t)$ and
$dW^{(2)}(t)$, with constant correlation parameter
$dW^{(1)}(t)dW^{(2)}(t)=\rho dt$. 
The drift coefficient for stochastic asset returns is given by the mean return of the asset where $\mu\in \mathbb{R}$ and the diffusion coefficient is given by $\sqrt{\sigma (t)}S(t)$. 

The drift coefficient of the asset variance is given by
$\kappa(\sigma(t))^\alpha(\widetilde{\theta}-\sigma(t))$,
where constants $\kappa \geq 0$ and $\theta
\geq 0$ are the mean reversion speed of $\sigma(t)$ and the long run
mean of $\sigma(t)$, respectively. The diffusion
coefficient is given by $v(\sigma(t))^\beta$ where constant $v\geq 0$
is the volatility of volatility. The
constant riskless interest rate is denoted by $r \geq 0.$ 
The constants $\alpha,\beta$ 
determine the stochastic volatility model used. 

The class of stochastic volatility models \eqref{eq:SVmodels} includes
a number of known stochastic volatility models:
The most prominent stochastic volatility model, the \textit{Heston
  model} \cite{Hes93} (also called \textit{square root (SQR) model}) specifies the variance by
$$d\sigma(t) = \kappa \left(\theta -\sigma(t) \right) {\rm d}t + v \sqrt{\sigma(t)} {\rm d}W^{(2)}(t).$$
Other known stochastic volatility models include the \textit{GARCH}
(or \textit{VAR model}) model, see \cite{Duan95}, where the stochastic variance is modelled by
$$d\sigma(t) = \kappa \left(\theta -\sigma(t) \right) {\rm d}t + v \sigma(t) {\rm d}W^{(2)}(t),$$
and the \textit{3/2 model} \cite{Lewis00} in which the variance follows the process
$$d\sigma(t) = \kappa \left(\theta -\sigma(t) \right) {\rm d}t + v \sigma^{\frac{3}{2}}(t) {\rm d}W^{(2)}(t).$$
All of the three  stochastic volatility models mentioned above use a linear mean-reverting drift for the stochastic process of the variance $v(t)$, 
but there are also models, in which the drift is mean reverting in a
non-linear fashion.
Following \cite{ChJaMi08}, we denote these models with an additional ``N'':
in the \textit{SQRN model} the stochastic variance follows
$$d\sigma(t) = \kappa \sigma(t)\left(\theta -\sigma(t) \right) {\rm d}t + v \sqrt{\sigma(t)} {\rm d}W^{(2)}(t) ,$$
in the \textit{VARN model}
$$d\sigma(t) = \kappa \sigma(t)\left(\theta -\sigma(t) \right) {\rm d}t + v \sigma(t) {\rm d}W^{(2)}(t) ,$$
and in  the \textit{$3/2$-N model}
$$d\sigma(t) = \kappa \sigma(t)\left(\theta -\sigma(t) \right) {\rm d}t + v \sigma^{\frac{3}{2}}(t) {\rm d}W^{(2)}(t) ,$$
see \cite{ChJaMi08}.

Applying standards arbitrage arguments and Ito’s lemma to the
class of stochastic volatility models \eqref{eq:SVmodels}, we can
derive the following second order partial differential equation for
any financial derivative $V(S,\sigma,t)$, to be solved backwards in
time with $0 < S < \infty $, $0 < \sigma < \infty$, $t \in [0,T)$:
\begin{equation}
\label{PDE}
V_{t} + \frac{S^2\sigma}{2}V_{SS} + \rho v \sigma^{\beta+\frac{1}{2}} S V_{S\sigma} + \frac{v^2\sigma^{2\beta}}{2}V_{\sigma\sigma} + rSV_{s} +[\kappa\sigma^{\alpha}(\theta  - \sigma) - \lambda_0\sigma]V_{\sigma} -rV = 0 .
 \end{equation}
Here, $\lambda_0\sigma(t)$ is the market price of volatility
risk, where $\lambda_0 \in \mathbb{R}$, which is usually assumed to be
proportional to the variance. In the following we assume $\lambda_0 =
0$ for streamlining the presentation. The generalisation to the case
$\lambda_0 \neq 0$ is straightforward by consistently adding in the
additional term in the coefficient of $V_\sigma$. The boundary conditions and final condition are
determined by the type of financial derivative $V(S,\sigma,t)$ we are
solving for. The boundary conditions of any European option will depend on a prescribed exercise price, denoted here by $E>0$. For example, in the case of the European Put Option:
\begin{align*}
    V(S,\sigma, T) &= \max(E-S,0),  & 0<&S<\infty, \; 0<\sigma<\infty,\\
    \lim_{S\to\infty}V(S,\sigma, t) &= 0, & 0<&\sigma<\infty, \; 0< t<T,\\
    V(0,\sigma, t) &= E\exp(-r(T-t)), & 0<&\sigma<\infty, \; 0< t<T,\\
    \lim_{\sigma\to\infty}V_{\sigma}(S,\sigma, t) &= 0, & 0<&S<\infty,\; 0< t<T,\\
\end{align*}
 The remaining boundary condition at $\sigma=0$ can be obtained by looking at
 the formal limit $\sigma\to 0$ in \eqref{PDE}, i.e.,
 \begin{equation}
   V_t+rSV_S+\kappa\theta V_\sigma-rV= 0,\quad T> t\geq 0,\;S>0,\; \text{as } \sigma\to  0.
   \label{boundary3}
 \end{equation}
 This boundary condition is used frequently, e.g.\ in \cite{IkoToi07,ZvFoVe98}.
 Alternatively, one can use a homogeneous Neumann condition
 \cite{ClaPar99}, i.e.,
 \begin{equation}
     V_{\sigma}(S,0,t) = 0, \quad 0<S<\infty, \; 0 < t<T.
 \end{equation}

\noindent By using a change of variables:
$$ x=\ln\frac{S}{E},\hskip10pt y = \frac{\sigma}{v}, \hskip10pt \tau=T-t, \hskip8pt u = \exp(r\tau)\frac{V}{E}    $$
we transform the partial differential equation to an convection-diffusion equation in two spatial dimensions with a mixed derivative term. The transformed partial differential equation and boundary/initial conditions are now satisfied by $u(x,y,\tau)$, where $x \in \mathbb{R}$, $y >0$, $\tau \in (0,T]$:
\begin{equation} \label{PDEtransf}
u_{\tau}=\frac{vy}{2}u_{xx}+\frac{(vy)^{2\beta}}{2}u_{yy} + \rho (v
y)^{\beta+\frac{1}{2}} u _{xy} + \Big(r-\frac{vy}{2}\Big)u_{x} +
\left(\kappa \left(vy\right)^{\alpha}\frac{\theta  - vy}{v}\right) u_y,
 \end{equation}
\begin{subequations}
\begin{align}
   u(x,y,0) &= \max(1-\exp(x),0), & -\infty<&x<\infty, 0<y<\infty, \\
    \lim_{x\to\infty}u(x,y,\tau) &= 0 , &  0<&y<\infty, 0\leq \tau<T ,\\
    \lim_{x\to -\infty}u(x,y,\tau) &= 1 , &   0<&y<\infty, 0\leq
                                                  \tau<T ,\\
\label{bcymax}
    \lim_{y\to \infty}u_{y}(x,y,\tau)&=0 , & -\infty<&x<\infty, 0<
                                                       \tau\leq T,\\
\label{bcymin}
    \lim_{y\to 0}u_{y}(x,y,\tau)&=0 , &   -\infty<&x<\infty, 0< \tau\leq T.
\end{align}
\end{subequations}
In order to discretise the problem and solve numerically, we truncate our spatial boundaries to finite values. Take $ L_{1} \leq x \leq K_{1} $, where $ L_{1} < K_{1}$, and $ L_{2} \leq y \leq K_{2} $, where $0<L_2<K_{2}$, so that the spatial domain forms a closed rectangle in $\mathbb{R}^2$ of $M \times N$ points with uniform spacing of $\Delta_{x}$ in the $x$-direction and $\Delta_{y}$ in the $y$-direction:
$$x_{i} = L_{1} +(i-1)\Delta_{x},\; i=1,2,\ldots, M, \hspace{20pt} y_{j} = L_2 + (j-1)\Delta_{y},\;  j=1,2,\ldots, N.$$
The lower $y$-boundary is truncated to $ L_{2}> 0$ to ensure
non-degeneracy of the partial differential equation for all values of
$y$. 
We assume cell aspect ratios to be moderate.
 We also take a uniform partition of $\tau\in [0,T]$ into $P$ points such that $\tau_{k} =  (k-1)\Delta_{\tau}$, where $k = 1,2,\ldots,P$. We denote the discrete approximation of $u((i-1)\Delta_x,(j-1)\Delta_y,(k-1)\Delta_\tau)$ by $u^{k}_{i,j}$ and $U^n=(u_{i,j}^n)_{i,j}$.

\section{Hundsdorfer-Verwer ADI splitting scheme}
\label{sec:HV}

\noindent We consider the Alternating Direction Implicit (ADI)
time-stepping numerical method proposed by Hundsdorfer and Verwer
\cite{Hund02, Verwer}. Our partial differential equation \eqref{PDEtransf} 
takes the form $u_{\tau} =F(u)$. We employ the splitting $F(u) =
F_{0}(u)+F_{1}(u)+F_{2}(u) $  where unidirectional and mixed
derivative differential operators are given by:
\begin{multline}\label{splitting}
F_{0}(u) = \rho (v y)^{\beta+\frac{1}{2}} u _{xy}, \; F_{1}(u) =
\frac{vy}{2}u_{xx}+\Big(r-\frac{vy}{2}\Big)u_{x}, \\ F_{2}(u)=\frac{(vy)^{2\beta}}{2}u_{yy}+
\left(\kappa \left(vy\right)^{\alpha}\frac{\theta  - vy}{v} \right)u_y .
\end{multline}
We consider \eqref{PDEtransf} with the splitting \eqref{splitting} 
and look for a semi-discrete approximation $U^n
\approx u(\tau_n)$  at time $n\Delta_{\tau}$.
Given an approximation $U^{n-1}$ we
can calculate an approximation for $U^{n}$ at time
$n\Delta_{\tau}$ using the differential operators from
\eqref{splitting}:
\begin{subequations}
\label{HVscheme}
\begin{align}
Y_0 &=  U^{n-1}+\Delta_{t}F(U^{n-1}),  \\
Y_1 &=   Y_0 + \phi \Delta_{t}(F_{1}(Y_1)-F_{1}(U^{n-1})), \\
Y_2 &=  Y_1 + \phi \Delta_{t}(F_{2}(Y_2)-F_{2}(U^{n-1})),\\
\widetilde{Y}_0 &= Y_0 + \psi \Delta_{t}(F(Y_2)-F(U^{n-1})), \\
\widetilde{Y}_1 &=  \widetilde{Y}_0+\phi \Delta_{t}(F_{1}(\widetilde{Y}_1)-F_{1}(Y_2)),\\
\widetilde{Y}_2 & =  \widetilde{Y}_1+\phi \Delta_{t}(F_{2}(\widetilde{Y}_2)-F_{2}
(Y_2)),\\
U^n &= \widetilde{Y}_2.
\end{align}
\end{subequations}
The parameter $\psi$ is taken to be $\psi=1/2$ to ensure
second-order accuracy in time. The parameter $\phi$ is typically fixed
to $\phi=1/2$. Larger values give 
stronger damping of the implicit terms while lower values return
better accuracy. The role of $\phi$ is discussed in
\cite{Hund02}. Its influence in the connection with high-order
spatial approximations is investigated numerically in
\cite{DM16}.

The first and fourth step in \eqref{HVscheme} can be solved explicitly, while the
remaining steps are solved implicitly. Our aim is to derive high-order
spatial discretisations of the differential operators. Following
\cite{DuFoRi13} we combine high-order compact finite difference
methods for the implicit steps with a (classical, non-compact)
high-order stencil for the explicit steps.


\section{High-order compact scheme for implicit steps}
\label{sec:HOC}

For $F_{1}(u)$, consider the one-dimensional convection-diffusion equation 
\begin{equation}
  \label{eq:rem1eqn}
   u_{xx} + c_1u_x = c_2g 
\end{equation}
with constants $c_1=2r/(vy)-1$ and $c_2=2/(vy)$. To discretise the partial derivatives
in \eqref{eq:rem1eqn}, we employ standard, centered second-order
finite difference operators, denoted by $\delta_{x0}$ and $\delta_x^2
$. The second-order terms in the truncation error involve
higher-order partial derivatives,
$u_{xxx}$ and $u_{xxxx}$.
Hence, if we can find second-order accurate expressions for $u_{xxx}$ and
$u_{xxxx}$, using only information on the compact stencil, then it will
be possible to approximate $u_x$ and $u_{xx}$ with fourth order
accuracy on the compact stencil. By differentiating \eqref{eq:rem1eqn}  once
and twice with respect to $x$, respectively, it is possible to express
$u_{xxx}$ and $u_{xxxx}$ in terms of first- and second-order
derivatives of $u$ and $g$ with respect to $x$.
We obtain the following relations, concisely written in matrix form,
$$\begin{bmatrix}
    1 & 0 & \frac{1}{6} & 0 \\
    0 & 1 & 0 & \frac{1}{12} \\
    0 & c_1\Delta_x^2 & 1 & 0 \\
  0 & 0 & c_1 & 1 
\end{bmatrix}\!\!
\begin{bmatrix}
  u_{x}   \\
  u_{xx}     \\
  \Delta_x^2 u_{xxx}     \\
  \Delta_x^2 u_{xxxx}     
\end{bmatrix}\! \!
= \!\!\begin{bmatrix}
    \delta_{x0} u_{i,j}    \\
    \delta_x^2 u_{i,j}     \\
    c_2\Delta_x^2 g_{x}      \\
    c_2\Delta_x^2g_{xx}     
  \end{bmatrix}
  +
\begin{bmatrix}\mathcal{O}(\Delta_x^4) \\ \mathcal{O}(\Delta_x^4) \\ 0 \\ 0 
\end{bmatrix}\! \!
=\! \!\begin{bmatrix}
    \delta_{x0} u_{i,j}    \\
    \delta_x^2 u_{i,j}     \\
   c_2 \Delta_x^2 \delta_{x0}g_{i,j}      \\
    c_2\Delta_x^2\delta_{x}^2g_{i,j}     
  \end{bmatrix}
  +\mathcal{O}(\Delta_x^4) .
  $$
This shows that only second-order approximations for $u_x,$ $u_{xx}$,
$g_x$ and $g_{xx}$ are needed. Using these relations to discretise
\eqref{eq:rem1eqn} and to replace the partial derivatives $u_{xxx}$ and
$u_{xxxx}$ in the truncation error, yields a fourth-order compact
approximation for \eqref{eq:rem1eqn} at all points of the
spatial grid except those that lie on the $x$- and $y$-boundaries. 
We refer to \cite{DM16} for more details of the derivation of the compact
high-order spatial discretisation.

To approximate $F_1(u)$ at points along the $x$ boundaries
of the inner grid of the spatial domain, we will require a
contribution from the Dirichlet values at the $x$-boundaries of the
spatial domain. We collect these separately in a vector
$d$. Details on the boundary conditions are given in
Section~\ref{sec:bc}.
The resulting linear system to be solved can be written in matrix form:
$$A_{x}{{u}} = B_{x}{{g}} + d,$$
where ${{u}} = (u_{2,2}, u_{2,3}, \ldots, u_{M-1,N-1})$,
${{g}} = (g_{2,2}, g_{2,3}, \ldots, g_{M-1,N-1})$. The coefficient matrices $A_x$ and $B_x$ are block diagonal matrices, with the following structure:
\begin{equation*} {A_x} = \left[ {\begin{array}{cccc}
A_x^{1,1}& 0 & 0  & 0\\
0 & A_x^{2,2} & 0 & 0\\
0 & 0 & \ddots & 0  \\
0 &0 &0 &A_x^{N-2,N-2} 
\end{array}} \right],\quad 
{B_x} = \left[ {\begin{array}{cccc}
B_x^{1,1}& 0 & 0  & 0\\
0 & B_x^{2,2} & 0 & 0\\
0 & 0 & \ddots & 0  \\
0 &0 &0 &B_x^{N-2,N-2} 
\end{array}} \right],
\end{equation*} 
where each $A_x^{j,j} =
\mathrm{diag}[a_{-1}^{j,j},a_0^{j,j},a_1^{j,j}]$ and $B_x^{j,j} =
\mathrm{diag}[b_{-1}^{j,j},b_0^{j,j},b_1^{j,j}]$ are tri-diagonal
matrices. Explicit expression for all coefficients are given in \cite{DM16}.

For $F_2(u)$ the derivation can be presented in a concise form, similar as
for $F_1(u)$, again we refer to \cite{DM16} for additional
details. Consider the one-dimensional convection-diffusion equation 
\begin{equation}
  \label{eq:rem2eqn}
  u_{yy} + c_1u_y = c_2g
\end{equation}
with $c_1(y)=2\kappa(vy)^{\alpha-2\beta}(\theta-vy)/v$ and $c_2(y)=2/(vy)^{2\beta}$, the necessary relations can
be concisely written in matrix form,
\begin{multline*}
\begin{bmatrix}
    1 & 0 & \frac{1}{6} & 0 \\
    0 & 1 & 0 & \frac{1}{12} \\
    c_{1}'\Delta_y^2 & c_{1}\Delta_y^2 & 1 & 0 \\
    c_{1}''\Delta_y^2 & 2c_{1}'\Delta_y^2 & c_{1} & 1 
  \end{bmatrix}
  \!\!
\begin{bmatrix}
  u_{y}   \\
  u_{yy}     \\
  \Delta_y^2 u_{yyy}     \\
  \Delta_y^2 u_{yyyy}     
\end{bmatrix}\\
 \!\!=  \!\!
\begin{bmatrix}
    \delta_{y0} u_{i,j}    \\
    \delta_y^2 u_{i,j}     \\
    \Delta_y^2(\delta_{y0}c_{2,j}g_{i,j} + c_{2,j}\delta_{y0}g_{i,j})     \\
    \Delta_y^2(\delta_{y}^2c_{2,j}g_{i,j} +
    2\delta_{y0}c_{2,j}\delta_{y0}g_{i,j} +
    c_{2,j}\delta_{y}^2g_{i,j}) 
  \end{bmatrix}
             +\mathcal{O}(\Delta_y^4) ,
\end{multline*}
where the first two lines of the system correspond to standard,
central second-order difference approximations, while the third and fourth are obtained from the repeated
differentiation of \eqref{eq:rem2eqn}. Using these relations to discretise
\eqref{eq:rem2eqn} and to replace the partial derivatives $u_{yyy}$ and
$u_{yyyy}$ in the truncation error, yields a fourth-order compact
approximation for \eqref{eq:rem2eqn}.

We obtain a linear system which can be represented in matrix form: $$A_{y}{{u}} = B_{y}{{g}}$$
where ${{u}} = (u_{2,2}, u_{2,3}, \ldots, u_{M-1,N-1})$,
${{g}} = (g_{2,2}, g_{2,3}, \ldots, g_{M-1,N-1})$. 
We do not impose any boundary conditions in $y$-direction, but
discretise the boundary grid points with the same scheme, and handle
resulting ghost points via extrapolation; details on the boundary
conditions are given in Section~\ref{sec:bc}. 
The coefficient matrices $A_y$ and $B_y$ are block tri-diagonal matrices with the following structures:
\begin{center}
\begin{eqnarray*}
\mathbf{A_y} = \left[\begin{array}{ccccc}
A_y^{1,1} & A_y^{1,2} & 0 & 0 & 0  \\
A_y^{2,1} & A_y^{2,2} & A_y^{2,3} & 0 & 0\\
0 & \ddots & \ddots & \ddots & 0  \\
0 & 0 & A_y^{N-3,N-4} & A_y^{N-3,N-3} & A_y^{N-3,N-2} \\ 0 & 0&0 & A_y^{N-2,N-3} & A_y^{N-2,N-2} 
\end{array}\right], \\ \mathbf{B_y} = \left[\begin{array}{ccccc}
B_y^{1,1} & B_y^{1,2} & 0 & 0 & 0  \\
B_y^{2,1} & B_y^{2,2} & B_y^{2,3} & 0 & 0\\
0 & \ddots & \ddots & \ddots & 0  \\
0 & 0 & B_y^{N-3,N-4} & B_y^{N-3,N-3} & B_y^{N-3,N-2} \\ 0 & 0&0 & B_y^{N-2,N-3} & B_y^{N-2,N-2} 
\end{array}\right],
\end{eqnarray*}
\end{center}
where each $A_y^{j,j} = \mathrm{diag}[a^{i,j}]$ and $B_y^{j,j} =
\mathrm{diag}[b^{i,j}]$ are diagonal matrices. Explicit expression for all coefficients are given in \cite{DM16}.

\section{High-order scheme for explicit steps}
\label{sec:HOexpl}

\noindent The first and fourth steps of the ADI scheme
\eqref{HVscheme} operate only on previous approximations to explicitly 
calculate an updated approximation. The differential operator in these
steps takes the form of the right hand side of \eqref{PDEtransf}. For
the mixed derivative term it seems not to be possible to
exploit the structure of the differential operator to
obtain a fourth-order approximation on a compact computational
stencil. Hence, in order to maintain fourth-order accuracy of the
scheme in the explicit steps of \eqref{HVscheme}, the derivatives in
each differential operator  $F_0$, $F_1$ and $F_2$ are approximated
using classical, fourth-order central difference operators which
operate on a larger $5\times 5$-stencil in the spatial domain. Here we use the shift operator defined by:
$$ s_x = e^{\Delta_x\delta_x} \text{  where  } (s_x u)_{i,j} = u_{i+1,j},\quad
s_y = e^{\Delta_y\delta_y} \text{  where  } (s_y u)_{i,j} = u_{i,j+1}.$$
For $F_{1}(u) = \frac{vy}{2}u_{xx}-(\frac{vy}{2}-r)u_{x}$, we have the following scheme: 
\begin{multline*}
\Big[ \frac{vy}{2} u_{xx}+\Big (r-\frac{vy}{2}\Big)u_{x}\Big]_{i,j} = \frac{vy_j}{2}\left( \frac{-s_x^{-2} + 16 s_x^{-1} - 30 + 16s_x - s_x^2}{12\Delta ^2_x}\right)u_{i,j} \\ + \left( r - \frac{vy_j}{2} \right) \left( \frac{s_x^{-2} -8s_x^{-1} + 8s_x -s_x^2}{12\Delta_x} \right)u_{i,j}  + \mathcal{O}(\Delta^4_x).
\end{multline*}
For $F_{2}(u)=\frac{(vy)^{2\beta}}{2}u_{yy}+\frac{\kappa(vy)^{\alpha}(\theta-vy)}{v}u_{y}$, we have:
\begin{multline*}
\Big[\frac{(vy)^{2\beta}}{2}u_{yy}+\frac{\kappa(vy)^{\alpha}(\theta-vy)}{v}u_y\Big]_{i,j}
\\= \frac{(vy_j)^{2\beta}}{2}\left( \frac{-s_y^{-2} + 16 s_y^{-1} - 30
    + 16s_y - s_y^2}{12\Delta ^2_y} \right) u_{i,j} \\ +  \frac{\kappa(vy_j)^{\alpha}(\theta-vy_j)}{v}\left( \frac{s_y^{-2} -8s_y^{-1} + 8s_y -s_y^2}{12\Delta_y} \right)u_{i,j}   + \mathcal{O}(\Delta^4_y ).
\end{multline*}
Finally, for the mixed derivative term $F_0 = \rho (v y)^{\beta+\frac{1}{2}}u_{xy} $, the following computational stencil is used:
\begin{multline*}
    \Big[\rho (v y)^{\beta+\frac{1}{2}} u_{xy}\Big]_{i,j} \\= \rho
    (vy_j)^{\beta + \frac{1}{2}}\left( \frac{s_x^{-2} -8s_x^{-1} +
        8s_x - s_x^{2}}{12\Delta_x}\right)\left( \frac{s_y^{-2}
        -8s_y^{-1} + 8s_y - s_y^{2}}{12\Delta_y} \right)u_{i,j}  \\
+ \mathcal{O}(\Delta^4_x\Delta^4_y) + \mathcal{O}(\Delta^4_x) + \mathcal{O}(\Delta^4_y).
\end{multline*}
Using these fourth-order approximations, the first and fourth step in
\eqref{HVscheme} can be computed directly. The values at the spatial
boundaries for each solution of the ADI scheme are determined by the
boundary conditions, the computational stencil is required for all
remaining points in the spatial domain. For the explicit steps, the
$5\times 5$-point computational stencil exceeds the spatial boundary
when we wish to approximate differential operator $F(u)$ at any point
along the boundary of the spatial domain's inner grid. For example if
we wish to evaluate $F(u_{2,2})$, we will require contributions from
ghost points which fall outside the spatial domain, as marked by
bullet points in Figure~\ref{fig:ghostpoints}.
\begin{figure}[t]
$$\begin{array}{c|ccccc}
\bullet & \mathrm{u_{4,1}} & u_{4,2} & u_{4,3} & u_{4,4} \\
\bullet & \mathrm{u_{3,1}} & u_{3,2} & u_{3,3} & u_{3,4} \\
\bullet & \mathrm{u_{2,1}} & {{u_{2,2}}} & u_{2,3} & u_{2,4} \\
\bullet & \mathrm{u_{1,1}} & \mathrm{u_{1,2}} & \mathrm{u_{1,3}} & \mathrm{u_{1,4}} \\
\hline 
\odot & \circ & \circ & \circ & \circ 
\end{array}$$
\caption{Example: evaluation of $F(u_{2,2})$ using the $5\times 5$-point computational stencil in the lower left
  corner of the computational domain; ghost points outside the
  computational domain at which values are
extrapolated from the interior of the domain are marked by bullets ($\bullet$,$\circ$,$\odot$),
grid points on the boundary are set in Roman.}
\label{fig:ghostpoints}
\end{figure}
We extrapolate information from grid points $u(x_i,y_j)$, where $i =
1,\ldots ,M-1,$ $j = 1,\ldots, N-1$ to establish values at these ghost
points for the purpose of evaluating the differential operator $F(u)$ at any point along the boundary of the inner grid of the spatial domain. To calculate the values at these ghost points, we use the following five-point extrapolation formulae for three cases:
\begin{align*}
& x=L_1 (\bullet)\text{ :} &u_{i,0} &= 5u_{i,1} - 10u_{i,2} + 10 u_{i,3} -5u_{i,4} + u_{i,5} + \mathcal{O}(\Delta^5_x) ,\\
& y=L_2 (\circ)\text{ :} &u_{0,j} &=  5u_{1,j} - 10u_{2,j} + 10 u_{3,j} -5u_{4,j} + u_{5,j} + \mathcal{O}(\Delta^5_y),\\
& x=L_1, y=L_2 (\odot)\text{ :} &u_{0,0} &=  5u_{1,1} - 10u_{2,2}+ 10 u_{3,3} -5u_{4,4} + u_{5,5} +\mathcal{O}(\Delta^5_x) \\& & &\hspace*{-1.5cm}  +\mathcal{O}(\Delta^4_x\Delta_y)
 +\mathcal{O}(\Delta^3_x\Delta^2_y) +\mathcal{O}(\Delta^2_x\Delta^3_y) +\mathcal{O}(\Delta_x\Delta^4_y) +\mathcal{O}(\Delta^5_y).
\end{align*}
The extrapolation at the $x=K_1$ and $y=K_2$ boundaries
and the remaining three corners is handled analogously. 

\section{Solving the high-order ADI scheme}
\label{sec:HOsolve}

\noindent Starting from a given $U^{n-1}$, the ADI
scheme \eqref{HVscheme} involves six approximation steps to obtain $U^{n}$, the
solution at the next time level. The first approximation $Y_0$ can be
solved for explicitly using the $5\times 5$-point computational
stencil derived in Section~\ref{sec:HOexpl}. The second approximation
for our solution, denoted by $Y_1$, has to be solved for implicitly:
\begin{align}
\label{stepximpl}
Y_1 =& Y_0 + \phi{\Delta_{t}} (F_{1}(Y_1)-F_{1}(U^{n-1}))\quad
\Longleftrightarrow \quad F_1(Y_1-U^{n-1})= \frac1{\phi\Delta_t}(Y_1-Y_0).
\end{align}
We apply the fourth-order compact scheme established in
Section~\ref{sec:HOC} to solve \eqref{stepximpl}. In matrix form we obtain
 $$A_x(Y_1-U^{n-1}) = B_x\Big(\frac{1}{\phi\Delta_t}(Y_1-Y_0)\Big)+ d.$$
Collecting unknown $Y_1$ terms on the left hand side and known terms
$Y_0$, $U^{n-1}$ and $d$ on the right hand side we get
$$\left(B_x-\phi\Delta_t A_x\right)Y_1 = B_{x}Y_0 - \phi\Delta_tA_{x}U^{n-1}- \phi\Delta_td.$$
To solve, we invert the tri-diagonal matrix
$\left(B_x-\phi\Delta_t A_x\right)$. 
For the third step of the ADI scheme, we proceed analogously, and use
the the high-order compact scheme presented in Section~\ref{sec:HOC}
to solve for $Y_2$ implicitly. The fourth, fifth and sixth step of the
ADI scheme are performed analogously as the first, second
and third steps, respectively.  

Note that the matrix $\left(B_x-\phi\Delta_t A_x\right)$ appears
twice in the scheme \eqref{HVscheme}, in the second and fifth step. Similarly,
$\left(B_y-\phi\Delta_tA_y\right)$ appears in the third and the sixth step. Hence, using LU-factorisation, only two matrix inversions are
necessary in each time step of scheme \eqref{HVscheme}.
Moreover, since the coefficients in the partial differential equation
\eqref{PDEtransf} do not depend on time, and the matrices are therefore
constant, they can be LU-factorised before iterating in time to obtain
a highly efficient algorithm.

The combination of the fourth-order spatial discretisation presented in
Section~\ref{sec:HOC} and \ref{sec:HOexpl} with the second-order time
splitting \eqref{HVscheme} yields a high-order ADI scheme with order of consistency two in time and four in space.

\section{Boundary conditions}
\label{sec:bc}

\noindent For the case of the Dirichlet conditions at $x=L_1$ and
$x=K_1$ we impose 
\begin{align*}
 u(L_1,y_j,\tau_k)&= 1-e^{r\tau+L_1}, & j&=1,2,\ldots,N,\;k=1,2,\ldots
, \\
u(K_1,y_j,\tau_k)&= 0, &  j&=1,2,\ldots,N,\;k=1,2,\ldots.
\end{align*}
Using the homogeneous Neumann conditions \eqref{bcymax} and \eqref{bcymin} which are correct in
the limit $y\to \infty$ and $y\to 0$, respectively, at the (finite)
boundaries $y=L_2>0$ and $y=K_2$ would result in a dominant error along
these boundaries. Hence, we do not impose any boundary condition at
these two boundaries but discretise the partial differential equation using the computational stencil
from the interior.
The values of the unknown on the boundaries are set by extrapolation
from values in the interior. This introduces a numerical error, and it
needs to be considered that the order of extrapolation should be high
enough not to affect the overall order of accuracy. We refer to Gustafsson \cite{GusBC} to discuss the influence of the  
order of the approximation on the global convergence rate.
We use the following extrapolation formulae:
\begin{align*}
& u^k_{i,1} = 5u^k_{i,2} -10u^k_{i,3} +10u^k_{i,4} -5u^k_{i,5}+u^k_{i,6} +\mathcal{O}(\Delta^6_y),\\ & u^k_{i,N} = 5u^k_{i,N-1} -10u^k_{i,N-2} +10u^k_{i,N-3} -5u^k_{i,N-4} +u^k_{i,N-5} +\mathcal{O}(\Delta^6_y).
\end{align*}

\section{Sparse grid combination technique}
\label{sec:ct}

\noindent Due to the ADI splitting and the compactness of the finite
difference discretisation in the implicit steps, the computational effort grows linearly with the number of unknowns, namely
$\mathcal{O}(N \cdot M)$. In the following we use the so-called \textit{sparse grid combination technique\/} to reduce the number of grid nodes and thus also the computational effort. 
Sparse grids go back to  Smolyak \cite{Smolyak}, who used them for numerical integration. 
Zenger \cite{Zenger_3}, Bungartz et al. \cite{Bungartz_3} and Schiekofer \cite{Schiekofer} extended his idea and applied sparse grids to solve PDEs with finite element, 
finite volume and finite difference methods. These methods in general require hierarchical, tree-like data structures, which makes
the data structure management more complicated than in the full grid case. With the help of the sparse grid combination technique \cite{Zenger_1} this problem can be overcome.
Here, full tensor-based solutions are linearly combined to construct
the sparse grid solution. This allows us to use standard full grid PDE solvers. Hence, this approach is very versatile 
and broadly applicable. Furthermore, each sub-solution can be computed independently, which makes it easily parallelisable.

The combination technique is based on the error splitting structure of the underlying numerical scheme. 
 Let the numerical solution of the HO-ADI scheme be given by $u_{\textbf{\textit{l}}}$ with multi-index $\textbf{\textit{l}}=  (l_1,l_2)$ and mesh widths $\Delta_x = 2^{-l_1}(K_1-L_1)$, 
 $\Delta_y = 2^{-l_2}(K_2-L_2)$. We assume that our numerical scheme satisfies an error splitting structure of the form
\[
 u - u_{\textbf{\textit{l}}} = \Delta_x^4 w_1( \Delta_x ) + \Delta_y^4 w_2( \Delta_y ) +\Delta_x^4 \Delta_y^4 w_{1,2}(\Delta_x,\Delta_y),
\]
with functions  $w_1, w_2, w_{1,2}$ bounded by some constant $C \in \mathbb{R}^+$. The mesh widths $\Delta_x$ and $\Delta_y$ are independent of one another. Since the error functions $w_1$ and $w_2$ only depend on either $\Delta_x$ or
$\Delta_y$, we can subtract two solutions with the same mesh width in one coordinate direction, such that the error term cancels out. Exploiting this idea further leads to the
combination technique
\begin{align}\label{eq:ct}
 u_n^s = \sum_{|\textbf{\textit{l}} |_1 = n+1} u_{\textbf{\textit{l}}} - \sum_{|\textbf{\textit{l}} |_1 = n} u_{\textbf{\textit{l}}}.
\end{align}
Applying the error splitting from above, the lower order terms cancel out and we obtain
\begin{multline*}
 u_n^s = u + 2^{-4(n+1)}R_1  w_1(2^{-(n+1)}R_1 ) + 2^{-2(n+1)(R_2)} w_2(2^{-(n+1)}R_2)\\
 + 2^{-4(n+1)}R_1R_2 \sum_{i=0}^{n+1} w_{1,2}(2^{-i}R_1,2^{-(n+1-i)}R_2 )\\
 - 2^{-4n}R_1R_2 \sum_{i=0}^{n}
 w_{1,2}(2^{-i}R_1,2^{-(n-i)}R_2 ),
\end{multline*}
where $R_1=K_1-L_1$ and $R_2=K_2-L_2$.
As $w_1,$ $w_2$ and $w_{1,2}$ are bounded by $C$ the pointwise error is given by
\[
 | u_n^s-u | = \mathcal{O}(n 2^{-4 n}),
\]
which is equivalent to 
\begin{align}\label{eq:error_ct}
|u_n^s - u| = \mathcal{O}(\Delta^4 \log_2 ( \Delta^{-1}))
\end{align}
 for $\Delta=2^{-n}$. We observe that the error of the sparse grid combination technique is deteriorated by
a factor of $\log_2 ( \Delta^{-1})$ compared to the fourth-order full grid solution.

\begin{figure}[h]
 \centering\includegraphics{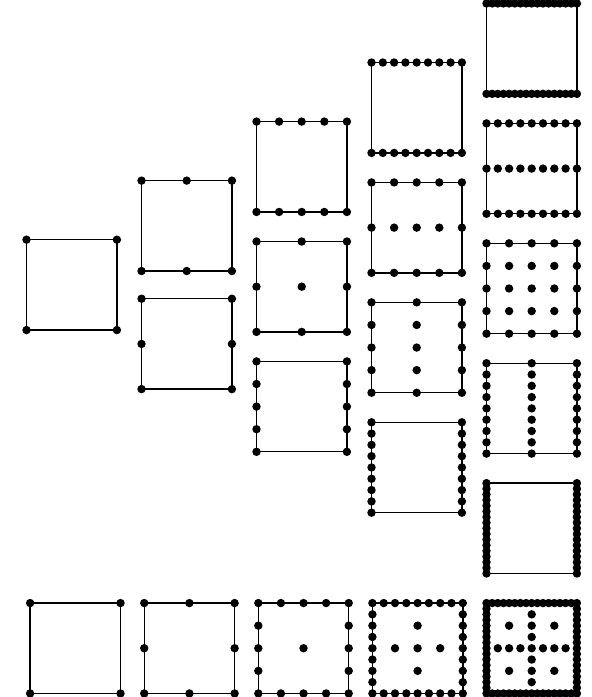}
  \caption{Sub-grids and sparse grid for $n=0,...,4$.}
  \label{fig:sparse_grid_hier}
 \end{figure}

Figure \ref{fig:sparse_grid_hier} shows the two-dimensional  grid hierarchy at levels $n=0,...,4$. The sparse grid in two dimensions at level $n$ consists of sub-grids, whose sum 
of refinement levels fulfils  $|\textbf{\textit{l}} |_1 = n$. Hence the number of grid points on each sub-grid grows with $\mathcal{O}(2^n)$. As the number of grids increases with $\mathcal{O}(n)$, 
this leads to $\mathcal{O}(n 2^n)$ nodes in the sparse grid. Let $\Delta=2^{-n}$, then this results in $\mathcal{O}(\Delta^{-1} \log_2(\Delta^{-1} ))$ grid points 
compared to $\mathcal{O}(\Delta^{-2})$ nodes in the full grid. Thus we
are able to reduce the number of grid nodes significantly while
maintaining a high accuracy.

It should be noted that for larger $n$ the combination technique as introduced above
involves solutions on grids which violate the assumption of moderate
cell aspect ratios which may lead to reduced accuracy and potential
instability of the scheme due to the extreme distortion of the grid. This
aspect of the combination technique is of general nature and not
specific to our scheme. A usual remedy would be to exclude solutions
on extremely distorted grids in \eqref{eq:ct}.
For further details we refer to the pertinent literature on sparse grids.

\section{Numerical experiments}
\label{sec:num}

\noindent In this section we test the proposed sparse grid high-order ADI
scheme. Beside the accuracy of the full grid solution we are also
interested in the efficiency of the combined sparse grid solution. 

It is well known that due to the non-smooth nature of the payoff
function in option pricing problems one cannot expect to see
higher-order in practice \cite{Forsyth}.
Some form of smoothing has to be applied to the initial condition. In
\cite{KrThWi70} suitable smoothing operators are identified in Fourier
space. Since the order of convergence of our high-order compact
scheme is four, we could use the smoothing operator $\Phi_4$
as in \cite{DuHe15}, given
by its Fourier transformation 
$$
\hat{\Phi}_4 (\omega)= \left(\frac{\sin\left( \frac{\omega}{2} \right) }{\frac{\omega}{2}}\right)^4 \left[ 1 + \frac{2}{3} \sin^2\left( \frac{\omega}{2} \right)\right].
$$
This leads to the smooth initial condition determined by
$$
\tilde{u}_0\left(x,y\right) =  \int\limits_{-3h}^{3h}
\int\limits_{-3h}^{3h}\Phi_4 \left(\frac{\tilde x}{h}\right)\Phi_4
\left(\frac{ \tilde y}{h}\right) u_0\left(x-\tilde x,y-\tilde y\right)\;
d\tilde x \,d\tilde y
$$
for any stepsize $h>0$, where $u_0$ is the original initial condition
and $\Phi_4(x)$ denotes the Fourier inverse of $\hat{\Phi}_4(\omega)$,
see \cite{KrThWi70}. As $h \rightarrow 0$, the smooth initial
condition $\tilde{u}_0$ tends towards the original initial condition
$u_0$ and the approximation of the smoothed problem tends towards the
true solution. For our numerical experiments
we use this smoothing operator which has already been applied  successfully to option pricing problems in \cite{DuHe15}. 

A numerical solution computed on a grid with $\Delta_x = \Delta \cdot (K_1-L_1)$, $\Delta_y = \Delta \cdot (K_2-L_2)$ and time step $\Delta_t = 5 \cdot \Delta^2$ serves as a reference solution, where 
$\Delta=2^{-8}$. Since the accuracy
of option prices close to the strike price is of highest interest from a practitioner's point of view, we compute the maximum absolute error in the region $[0.5 E, 2E] \times [0.05, 1]$. 
The grid parameters of the computational domain are chosen to be $L_1=-5$, $K_1 = 1.5$, $L_2=0.05$ and $K_2=2.5$. The parameters of the ADI method are $\psi=1/2$ and $\phi=1/2$, 
cf.\ Section~\ref{sec:HV}. The full grid solution is computed with step sizes $\Delta_x = \Delta \cdot (K_1-L_1)$, $\Delta_y = \Delta \cdot (K_2-L_2)$ and $\Delta_t = 5 \cdot \Delta^2$ with $\Delta = 2^{-n}$, 
while the sparse grid solution  $u_n^s$
is constructed according to definition (\ref{eq:ct}). In order to avoid instabilities due to the extreme distortion of the grid we neglect grids within the combination technique, 
where $l_i \leq 2$ for $i=1,2$. Thus, the finest resolution in one of the sub-grids along one coordinate direction is given by $\Delta = 2^{-(n-3)}$.

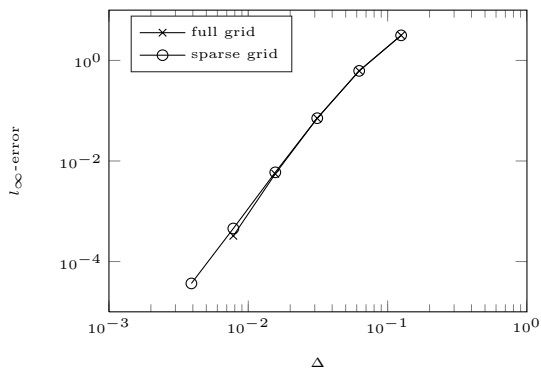
\begin{figure}[]
 \centering
   {\tiny{
%
%
\begin{tikzpicture}

\begin{axis}[%
width=5.5cm,
height=4cm,
at={(0.758in,0.481in)},
scale only axis,
xmode=log,
xmin=0.001,
xmax=1,
xminorticks=true,
xlabel = {$\Delta$},
ymode=log,
ymin=0.00001,
ymax=10,
yminorticks=true,
ylabel = {$l_{\infty}$-error},
axis background/.style={fill=white},
legend style={at={(0.052,0.794)},anchor=south west,legend cell align=left,align=left,draw=white!15!black}
]
\addplot [color=black,solid,mark=x,mark options={solid}]
  table[row sep=crcr]{%
0.125	3.15149485029395\\
0.0625	0.619064333669741\\
0.03125	0.0704535474559247\\
0.015625	0.00554603683681165\\
0.0078125	0.000324458133565741\\
};
\addlegendentry{full grid};
\addplot [color=black,solid,mark=o,mark options={solid}]
  table[row sep=crcr]{%
0.125	3.15149485029395\\
0.0625	0.619575690265831\\
0.03125	0.0709805801286798\\
0.015625	0.00594552641523549\\
0.0078125	0.000454689315144918\\
0.00390625	3.66822354749274e-05\\
};
\addlegendentry{sparse grid}
\end{axis}
\end{tikzpicture}%
}}
   \caption{Error decay of the full grid for $n=3,4,\ldots,7$ and sparse grid combination technique for $n=6,7,\ldots,11$. }
   \label{fig:experiment1}
 \end{figure}

We compare the performance of the high-order ADI scheme in the full and sparse grid case for a European put option with the parameters given in Table~\ref{tab:para}.
Figure~\ref{fig:experiment1} shows the maximum error plotted versus the
grid  resolution $\Delta$ for both cases. The
fourth-order compact finite difference scheme achieves an estimated
numerical convergence order of $3.33$, the error 
of the sparse grid solution decays slightly slower due to the logarithmic
factor in (\ref{eq:error_ct}). 

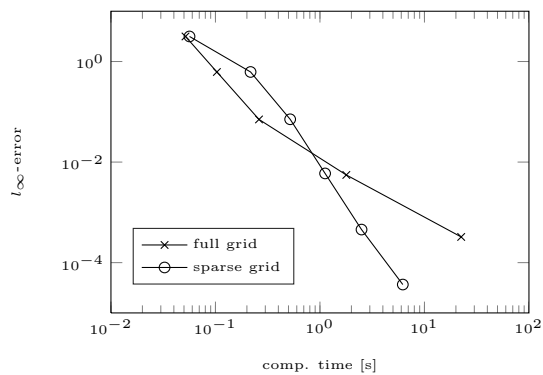
\begin{figure}[]
 \centering
   {\tiny{
%
%
\begin{tikzpicture}

\begin{axis}[%
width=5.5cm,
height=4cm,
at={(0.758in,0.481in)},
scale only axis,
xmode=log,
xmin=0.01,
xmax=100,
xminorticks=true,
xlabel={comp. time [s]},
ymode=log,
ymin=0.00001,
ymax=10,
yminorticks=true,
ylabel = {$l_\infty$-error},
axis background/.style={fill=white},
legend style={at={(0.052,0.094)},anchor=south west,legend cell align=left,align=left,draw=white!15!black}
]
\addplot [color=black,solid,mark=x,mark options={solid}]
  table[row sep=crcr]{%
0.051685	3.15149485029395\\
0.103235	0.619064333669741\\
0.261134	0.0704535474559247\\
1.788116	0.00554603683681165\\
22.469034	0.000324458133565741\\
};
\addlegendentry{full grid};
\addplot [color=black,solid,mark=o,mark options={solid}]
  table[row sep=crcr]{%
0.056458	3.15149485029395\\
0.216814	0.619575690265831\\
0.518752	0.0709805801286798\\
1.120118	0.00594552641523549\\
2.500236	0.000454689315144918\\
6.208251	3.66822354749274e-05\\
};
\addlegendentry{sparse grid}
\end{axis}
\end{tikzpicture}%
}} 
   \caption{Error versus computational time for the full grid for $n=3,4,\ldots,7$ and sparse grid combination technique for $n=6,7,\ldots,11$. }
   \label{fig:experiment2}
 \end{figure}

To illustrate the computational
efficiency we compare the run-time to the
accuracy in Figure \ref{fig:experiment2} for both approaches. 
We confirm that, as the mesh width decreases, the lower number of
employed grid nodes in the sparse grid method outweighs its slightly lower
convergence rate.  The serial implementation of the combination 
technique outperforms the full grid solver in the high accuracy
region, reducing the computational time by about an order of magnitude, while achieving a similar accuracy.

\begin{figure}
\begin{center}
\begin{tabular}{ll} \hline
Parameter & Value \\ \hline
Strike price & $E=100$ \\
Time to maturity & $T=1$\\
Interest rate & $r=0.05$\\
Volatility of volatility & $v=0.1$ \\
Mean reversion speed  & $\kappa=2$\\
Long run mean of volatility & $\theta=0.1$\\
Correlation & $\rho=-0.5$\\
Stochastic volatility drift parameter & $\alpha=0.5$\\
Stochastic volatility diffusion parameter & $\beta=0.5$\\ \hline
\end{tabular}
\caption{Parameters used in the numerical experiments.}
\label{tab:para}
\end{center}
\end{figure}

\subsection*{Acknowledgement}
\noindent 
BD acknowledges support by the Leverhulme Trust research project grant
`Novel discretisations for higher-order nonlinear PDE' (RPG-2015-69).
CH is supported by the
European Union in the FP7-PEOPLE-2012-ITN Program under
Grant Agreement Number 304617 (FP7 Marie Curie Action, Project Multi-ITN STRIKE - Novel Methods in Computational Finance).
Further CH acknowledges partial support from the bilateral German-Spanish Project
HiPeCa -- High Performance Calibration and Computation in Finance, Programme Acciones Conjuntas
Hispano-Alemanas financed by DAAD.
JM has been supported in part by a studentship under the EPSRC
Doctoral Training Partnership (DTP) scheme (grant number
EP/L505109/1).



\begin{thebibliography}{99}

\bibitem{BeamWarming} R.M.~Beam and R.F.~Warming, 
Alternating direction implicit methods for parabolic equations with a mixed derivative, 
{\em SIAM J.~Sci.~Stat.~Comput.} \textbf{1}(1), 1980.

\bibitem{BeGoMi10} E.~Benhamou, E.~Gobet and M.~Miri. Time dependent
  Heston model, {\em SIAM J.\ Finan.\ Math.} \textbf{1}, 289--325, 2010.

  \bibitem{Bungartz_3}
H.~Bungartz and M.~Griebel.
\newblock {Sparse grids}, {\em Acta~Numer.} \textbf{13}, 1--123, 2004.
  
\bibitem{ChJaMi08}
P.~Christoffersen, K.~Jacobs and K.~Mimouni.
\newblock Volatility dynamics for the {S}\&{P}500: evidence from realized volatility,
  daily returns, and option prices.
\newblock {\em Review of Financial Studies} \textbf{23}, 3141--3189, 2010.

\bibitem{ClaPar99} N.~Clarke and K.~Parrott. Multigrid for American
  option pricing with stochastic volatility. {\em Appl.\ Math.\ Finance}
  \textbf{6}(3), 177--195, 1999.

\bibitem{Douglas} J.~Douglas, Alternating direction methods for three
  space variables, \textit{Numer. Math.}, \textbf{4}, 41--63, 1962.

\bibitem{Gunn} J.~Douglas and J.E.~Gunn, A general formulation of alternating direction methods. I. Parabolic and hyperbolic problems, \textit{Numer.~Math.} \textbf{6}, 428--453, 1964.

\bibitem{Duan95}
J.~Duan.
\newblock The {GARCH} option pricing model.
\newblock {\em Math. Finance} \textbf{5}(1), 13--32, 1995.

\bibitem{DuFo12}
B.~D{\"u}ring and M.~Fourni{\'e}.
\newblock High-order compact finite difference scheme for option pricing in
  stochastic volatility models.
\newblock {\em J. Comput. Appl. Math.} \textbf{236}(17), 4462--4473, 2012.

\bibitem{DuFoHe14}
B.~D{\"u}ring, M.~Fourni{\'e} and C.~Heuer.
\newblock High-order compact finite difference schemes for option pricing in
  stochastic volatility models on non-uniform grids.
\newblock {\em J. Comput. Appl. Math.} \textbf{271}(18), 247--266, 2014.

\bibitem{DuFoJu04} B.~D\"uring, M.~Fourni\'e and A.~J\"ungel. Convergence
  of a high-order compact finite difference scheme for a nonlinear
  Black-Scholes equation. {\em Math.~Mod.~Num.~Anal.} \textbf{38}(2),
  359--369, 2004.

\bibitem{DuFoJu03} B.~D\"uring, M.~Fourni\'e and A.~J\"ungel. High-order
  compact finite difference schemes for a nonlinear Black-Scholes
  equation. {\em Intern.~J.~Theor.~Appl.~Finance} \textbf{6}(7),
  767--789, 2003.

\bibitem{DuFoRi13}
B.~D\"uring, M.~Fourni\'e and A.~Rigal.
High-order ADI schemes for convection-diffusion equations with mixed derivative terms.
In: {\em Spectral and High Order Methods for Partial Differential Equations - ICOSAHOM'12}, M.~Azaïez et al.\ (eds.), pp. 217--226, Lecture Notes in Computational Science and Engineering 95, Springer, Berlin, Heidelberg, 2013.

\bibitem{DuHe15} B.~D{\"u}ring and C.~Heuer.
\newblock High-order compact schemes for parabolic problems with mixed
  derivatives in multiple space dimensions.
\newblock {\em SIAM J.~Numer.~Anal.} \textbf{53}(5), 2113--2134, 2015.

\bibitem{Due09} B.~D\"uring. Asset pricing under information with
  stochastic volatility. {\it Rev.~Deriv.~Res.} \textbf{12}(2),
  141--167, 2009.

\bibitem{DM16} B.~D\"uring and J.~Miles. High-order ADI scheme for
  option pricing in stochastic volatility
  models. {\em J.\ Comput.\ Appl.\ Math.}, 2016,  \url{http://dx.doi.org/10.1016/j.cam.2016.09.040}. 

\bibitem{FournieRigal} M.~Fourni\'e and A.~Rigal.
High order compact schemes in projection methods for incompressible
viscous flows, {\em Commun.\ Comput.\ Phys.} \textbf{9}(4), 994--1019, 2011.
 
 \bibitem{Zenger_1}
M.~Griebel, M.~Schneider and C.~Zenger.
\newblock {A Combination Technique for the Solution of Sparse Grid Problems}.
\newblock {\em IMACS Elsevier, Iterative Methods in Linear Algebra},
  16:263--281, 1992.

\bibitem{Gupta} M.M.~Gupta, R.P.~Manohar and J.W. Stephenson, A single
  cell high-order scheme for the convection-diffusion equation with
  variable coefficients, \textit{Int. J. Numer. Methods Fluids},
  \textbf{4}, 641--651, 1984.
  
\bibitem{GusBC} B.~Gustafsson. The convergence rate for difference
  approximation to general mixed initial-boundary value problems. {\em
    SIAM J.\ Numer.\ Anal.} \textbf{18}(2), 179--190, 1981.

\bibitem{HeEhGu15} C.~Hendricks, M.~Ehrhardt and M.~G{\"u}nther,
  High-order ADI schemes for diffusion equations with mixed
  derivatives in the combination technique, 
{\em Appl.~Numer.~Math.} \textbf{101}, 36–-52, 2016.

\bibitem{Hes93} S.L.~Heston. A closed-form solution for options with
  stochastic volatility with applications to bond and currency
  options. {\it Review of Financial Studies} \textbf{6}(2), 327--343, 1993.
  
\bibitem{HouFou10} K.J.~in't~Hout and S.~Foulon. ADI finite difference
  schemes for option pricing in the Heston model with correlation.
  {\em Int.\ J.\ Numer.\ Anal.\ Mod.} \textbf{7}, 303--320, 2010.
  
\bibitem{HouWel07} K.J.~in’t~Hout and B.D.~Welfert. Stability of ADI
      schemes applied to convection-diffusion equations with mixed
      derivative terms. {\em Appl.\ Num.\ Math.\ } \textbf{57}, 19--35, 2007.

\bibitem{Verwer} W.~Hundsdorfer and J.G.~Verwer, Numerical solution of
  time-dependent advection-diffusion-reaction equations, Springer
  Series in Computational Mathematics, \textbf{33}, Springer-Verlag,
  Berlin, 2003.

\bibitem{Hund02} W.~Hundsdorfer, Accuracy and stability of splitting
  with stabilizing corrections, \textit{Appl.~Num.~Math.}
  \textbf{42}, 213--233, 2002.

\bibitem{HiMaSc05} N.~Hilber, A.~Matache and C.~Schwab. Sparse
  wavelet methods for option pricing under stochastic volatility.
{\em J.~Comput.~Financ.} \textbf{8}(4), 1--42, 2005.

\bibitem{IkoToi07} S.~Ikonen and J.~Toivanen. Efficient numerical methods
  for pricing American options under stochastic volatility.
  {\em Numer.~Methods Partial Differential Equations} \textbf{24}(1), 104--126, 2008.

\bibitem{KrThWi70} H.O.~Kreiss, V.~Thom{\'e}e and O.~Widlund. Smoothing
  of initial data and rates of convergence for parabolic difference
  equations. {\em Comm.\ Pure Appl.\ Math.} \textbf{23}, 241--259,
  1970.

\bibitem{LiaKha09} W.~Liao and A.Q.M.~Khaliq. High-order compact scheme
  for solving nonlinear Black-Scholes equation with transaction cost. {\em
    Int.\ J.\ Comput.\ Math.} \textbf{86}(6),
1009--1023, 2009. 

\bibitem{LaBlVe01} D.~Lanser, J,~Blom and J.~Verwer. Time integration
  of the shallow water equations in spherical geometry. {\em J.\
    Comp.\ Phys.} \textbf{171}, 373--393, 2001.

\bibitem{Lewis00}
A.L.~Lewis. {\em Option valuation under stochastic volatility}.
Finance Press, Newport Beach, CA, 2000.    

\bibitem{Fairweather} A.R.~Mitchell and G.~Fairweather,
  Improved forms of the alternating direction methods of Douglas,
  Peaceman, and Rachford for solving parabolic and elliptic equations,
  \textit{Numer.\ Math.\ } \textbf{6}, 285--292, 1964.

\bibitem{PeacmanRachford95} D.W.~Peaceman and H.H.~Rachford Jr., The numerical solution of parabolic and elliptic differential
equations, \textit{J.\ Soc.\ Ind.\ Appl.\ Math.\ } \textbf{3},  28--41, 1959.

\bibitem{Forsyth}
D.~M. Pooley, K.~R. Vetzal and P.~A. Forsyth.
\newblock {Convergence remedies for non-smooth payoffs in option pricing}.
\newblock {\em J.~Comput.~Financ.} \textbf{6}(4), 25--40, 2003.

\bibitem{Oosterlee} M.J.~Ruijter and C.W.~Oosterlee. Two-dimensional
  Fourier cosine series expansion method for pricing financial
  options, \textit{SIAM J. Sci. Comp.} \textbf{34}(5), 642--671, 2012.
  
\bibitem{Rigal99} A.~Rigal, Sch\'emas compacts d'ordre \' elev\'e: application aux probl\`emes bidimensionnels de diffusion-convection
instationnaire I, \textit{C.R.\ Acad.\ Sci.\ Paris.\ Sr.\ I Math.} \textbf{328}, 535--538, 1999.

\bibitem{Schiekofer}
T.~Schiekofer, {\em Die Methode der Finiten Differenzen auf d\"{u}nnen Gittern zur
  L\"{o}sung elliptischer und parabolischer partieller
  Differentialgleichungen}, PhD thesis, Universit\"at Bonn, 1999.

\bibitem{Smolyak}
S.~Smolyak, {Quadrature and interpolation formulas for tensor products of certain classes of functions},
{\em Dokl.\ Akad.\ Nauk SSSR} \textbf{148}, 1042--1045, 1963.

\bibitem{SpotzCarey} W.F.~Spotz and C.F.~Carey. 
  Extension of high-order compact schemes to time-dependent problems. 
  {\em Numer.\ Methods Partial Differential Equations} 
  \textbf{17}(6), 657--672, 2001.
  


\bibitem{TavRan00} D.~Tavella and C.~Randall.
   \emph{Pricing financial instruments: the finite difference method}.
   John Wiley \& Sons, 2000.
   
\bibitem{Zenger_3}
C.~Zenger.
\newblock {Sparse grids}.
\newblock Technical report, Institut f{\"{u}}r Informatik, Technische
  Universit{\"{a}}t M{\"{u}}nchen, Oct. 1990.
 
\bibitem{ZhuKop10} W.~Zhu and D.A.~Kopriva. A spectral element
  approximation to price European options with one asset and
  stochastic volatility. {\em J.~Sci.~Comput.} \textbf{42}(3),
  426--446, 2010.
  
\bibitem{ZvFoVe98} R.~Zvan, P.A.~Forsyth and K.R.~Vetzal. Penalty methods
  for American options with stochastic volatility. {\em
    J.~Comp.~Appl.~Math.} \textbf{91}(2), 199--218, 1998.
  
\end{thebibliography}
\end{document}